\documentclass[conference]{IEEEtran}
\usepackage{graphicx} 
\usepackage{float}
\usepackage{url}
\usepackage{siunitx}
\usepackage{cite}
\usepackage{amsmath,amssymb}
\usepackage{graphicx}
\usepackage{booktabs}
\usepackage{array}
\usepackage{makecell}
\usepackage{url}
\usepackage{xcolor}
\newcommand{\fcrctrl}{\mathrm{FCR}_{\mathrm{ctrl}}}
\newcommand{\fcramb}{\mathrm{FCR}_{\mathrm{amb}}}
\title{Controller-Only False Confirmation in Passive RF UAV Link Detection}
\author{
\IEEEauthorblockN{
Rajendra Upadhyay, Rajendra Paudyal, Al Nahian Bin Emran, Duminda Wijesekera
}
\IEEEauthorblockA{
Mason Innovation Labs, George Mason University, 3434 Washington Blvd, Arlington, VA 22201, USA\\
\{rupadhya, rpaudyal, abinemra, dwijesek\}@gmu.edu
}
}
\date{May 2026}

\begin{document}

\maketitle
\begin{abstract}
Passive radio frequency (RF) sensing is widely used for
counter-unmanned-aerial-vehicle (counter-UAV) detection. Existing studies commonly report high accuracy against background RF or Wi-Fi/Bluetooth interference, but rarely isolate controller-only operation without a linked aircraft. We present a dual-band software-defined radio (SDR) measurement study with ambient, controller-only, and linked states for three commercial
UAV platforms (DJI Phantom~3 4K, Hubsan H501S, DJI Mavic~Mini). Two USRP~B210 receivers simultaneously scan eight 2.4\,GHz and twelve 5.8\,GHz observation windows across twenty rounds. We first train an energy-based detector using linked and ambient scans only. At four ranked concurrent dwell steps (8\,s), it achieves 0.992 linked-versus-ambient balanced accuracy but false-confirms
30 out of 60 controller-only scans ($\fcrctrl=0.500$). We then include controller-only scans in training and use confirm, reject, and defer outputs under an explicit controller-only false-confirmation-rate (FCR) constraint. For the pooled all-platform analysis, the constraint reduces observed $\fcrctrl$ from 0.350 to 0.050, while confirm-linked TPR decreases from 0.900
to 0.400 and 42.1\% of scans are deferred. The corresponding compact scan performs substantially better for Hubsan and Mavic than for Phantom. A hardware-in-loop experiment reduces measured wall-clock time from 82.9\,s to 28.9\,s. These measurements show that linked-versus-background accuracy does not measure controller-only
false confirmation and that this error should be reported separately.
\end{abstract}

\begin{IEEEkeywords}
Counter-UAV, RF sensing, software-defined radio, USRP, passive detection, false confirmation, dual-band scanning.
\end{IEEEkeywords}

\section{Introduction}
\label{sec:intro}

Commercial unmanned aerial vehicle (UAV) systems use radio frequency (RF) links in unlicensed Industrial, Scientific, and Medical (ISM) bands for command, control, telemetry, and live video capture. Among different sensing modalities, passive RF sensing is an attractive approach for counter-UAV monitoring, as it does not require active transmission and can operate continuously over the same bands used by the target system~\cite{ezuma2019,alsad2019,seidaliyeva2023}. Beyond standalone passive RF sensing, complementary approaches have also used cellular network measurements, including handover behavior and received-signal variation, to detect uncooperative UAVs from radio-access-network observations~\cite{upadhyaybordersecurity}.

A practical counter-UAV system must distinguish at least three RF conditions. The first is ambient activity: no controller or aircraft is present, and any RF in the band is from other emitters such as Wi-Fi. The second is controller-only: a UAV controller is powered and may be beaconing or transmitting, but the aircraft is off. The third is a controller linked to a UAV: the aircraft is powered, paired, and active, and command, telemetry, and possibly live video traffic flow between the controller and aircraft. If the drone-detection system merges the second and third cases, it produces controller-only false confirmation, where a linked UAV alarm is raised when no aircraft is present.

This work analyzes the measurement gap in which controller-only false confirmation is not accounted separately in RF-based UAV detection. Most studies treat background/no-drone RF, Wi-Fi/Bluetooth interference~\cite{ezuma2019} or aggregate RF interferers~\cite{rss2019} as negative conditions, but do not isolate a powered UAV controller with no linked aircraft as a distinct negative class. In our measurements, this case corresponds to the controller powered on while the aircraft remains off, with the phone condition documented for each protocol. We test the consequence of this omission directly. An energy-feature detector trained only on linked and ambient scans reaches 0.992 balanced accuracy on those two states, but false-confirms 30 of 60 controller-only scans that were excluded from training. This does not establish the behavior of every published detector but it shows that linked-vs-background accuracy alone cannot measure this error. We address the gap with a controlled dual-band SDR measurement study and a constrained decision protocol. Our contributions are:

\begin{itemize}
    \item A dual-band SDR measurement dataset across three commercial UAV platforms (DJI Phantom~3 4K, Hubsan H501S, DJI Mavic~Mini) with explicit ambient, controller-only, and linked states across twenty collection rounds, and no-phone and phone-connected Phantom protocols for measuring the effect of phone-connected live-view configuration.
    
    \item An omitted-negative audit and a controller-only FCR-constrained decision framework with confirm, reject, and defer outputs under leave-one-round-out (LORO) evaluation.
    
    \item A per-platform characterization showing low controller-only FCR for Hubsan and Mavic Mini at the selected operating point, and weaker linked/controller separation for Phantom in these measurements.
    \item A hardware-in-the-loop measurement of wall-clock scan time for full and compact ranked scans on a dual-USRP B210 setup.
\end{itemize}

The code, analysis scripts and processed dataset used in this study are available at https://github.com/rajyay/controller-only-rf-uav.
\section{Related Work}
  \label{sec:related}

  \subsection{RF-based UAV datasets and benchmarks.}
  DroneRF is an early public RF-UAV dataset with recordings from three drones in various operating modes like on-and-connected, hovering, flying, and video recording, together with background RF recordings
  with no drones present~\cite{allahham2019dronerf,alsad2019}. RFUAV expands the scale of benchmarking with approximately 1.3\,TB of raw RF data collected from 37 UAVs using USRP hardware, and provides
  preprocessing and evaluation tools for detection and identification across SNR conditions~\cite{shi2025rfuav}. CageDroneRF is a recent large-scale benchmark with real-world and controlled RF-cage captures, 23 UAV
  models, remote-controller signals, operational variants, no-drone recordings, and Wi-Fi interferer recordings~\cite{rostami2026cagedrone}.

  \subsection{Interference-aware RF detection.}
  Ezuma et al. study passive RF detection and classification of UAV controller signals in the presence of Wi-Fi and Bluetooth interference. Their multistage detector separates UAV-controller signals from background
  noise and interference, and then classifies UAV controllers using RF fingerprints~\cite{ezuma2019}. Medaiyese et al. propose a hierarchical RF-UAV framework using the CardRF dataset, which includes UAV signals, UAV-
  control signals, Bluetooth devices, and Wi-Fi devices~\cite{medaiyese2022hierarchical}.

  \subsection{Multi-class RF-UAV learning.}
  Other RF-UAV studies focus on learned features and classification across drone or controller types using spectral, time-frequency, convolutional, or residual-network models~\cite{basak2021,deep2d2023,robust2024}.
  SDR-based monitoring systems for drone RF signatures have also been demonstrated using wideband or real-time RF sensing pipelines~\cite{sdrrt2026,wideband2024}.

  \subsection{Spectrum sensing and scheduling.}
  Wideband spectrum sensing and scheduling have also been studied for UAV networks and unmanned traffic-management settings~\cite{collab2023,fed2024}. That line of work is related in its treatment of wideband sensing
  and scheduling, but the UAV is generally part of the sensing or networked system rather than the passive RF target being confirmed.

  \textbf{Comparison with other studies.}
  Across the works above, negative conditions are typically background/no-drone RF, Wi-Fi/Bluetooth interference, other same-band activity, or non-target classes in a multi-class identification problem. A powered UAV
  controller with no linked aircraft is rarely isolated as a labeled negative condition against linked-UAV positives. Table~\ref{tab:related} summarizes this distinction. This paper evaluates controller-only false
  confirmation as a separate error mode and reports linked-UAV confirmation under a controller-only FCR constraint. The cited public datasets do not provide the same protocol-matched linked and powered-controller-with-aircraft-off labels, so they cannot be used for a direct evaluation of this FCR.
  
  \begin{table}[h]
    \centering
    \caption{Comparison with earlier literature. ``Yes'' indicates presence of signal class. ``n.r'' indicates class is not reported. ``Target'' denotes controller or UAV-control RF treated as a signal class, not as a aircraft-off negative.}
    \label{tab:related}
    \setlength{\tabcolsep}{2pt}
    \renewcommand{\arraystretch}{1.08}
    \resizebox{\columnwidth}{!}{%
    \begin{tabular}{l c c c c c}
    \toprule
    Work &
    Ambient &
    \makecell{Wi-Fi/BT/\\interference} &
    \makecell{Multiple UAV/\\controller types} &
    \makecell{Controller-only\\negative} &
    \makecell{Linked vs.\\controller-only} \\
    \midrule
    DroneRF~\cite{allahham2019dronerf}                & Yes & n.r  & Yes & n.r & n.r \\
    Ezuma et al.~\cite{ezuma2019}                     & Yes & Yes & Yes & Target & n.r \\
    Medaiyese et al.~\cite{medaiyese2022hierarchical} & n.r  & Yes & Yes & Target & n.r \\
    RFUAV~\cite{shi2025rfuav}                         & n.r  & n.r  & Yes & n.r & n.r \\
    CageDroneRF~\cite{rostami2026cagedrone}           & Yes & Yes & Yes & Target & n.r \\
    \textbf{This work}                                & Yes & env. & Yes & \textbf{Yes} & \textbf{Yes} \\
    \bottomrule
    \end{tabular}%
    }
  \end{table}

\section{Measurement System and Dataset}\label{sec:system}
\label{sec:measurement+Data}

\subsection{Hardware and Capture Configuration}
\label{ssec:configs}

\begin{figure}[h]
    \centering
    \includegraphics[width=\linewidth]{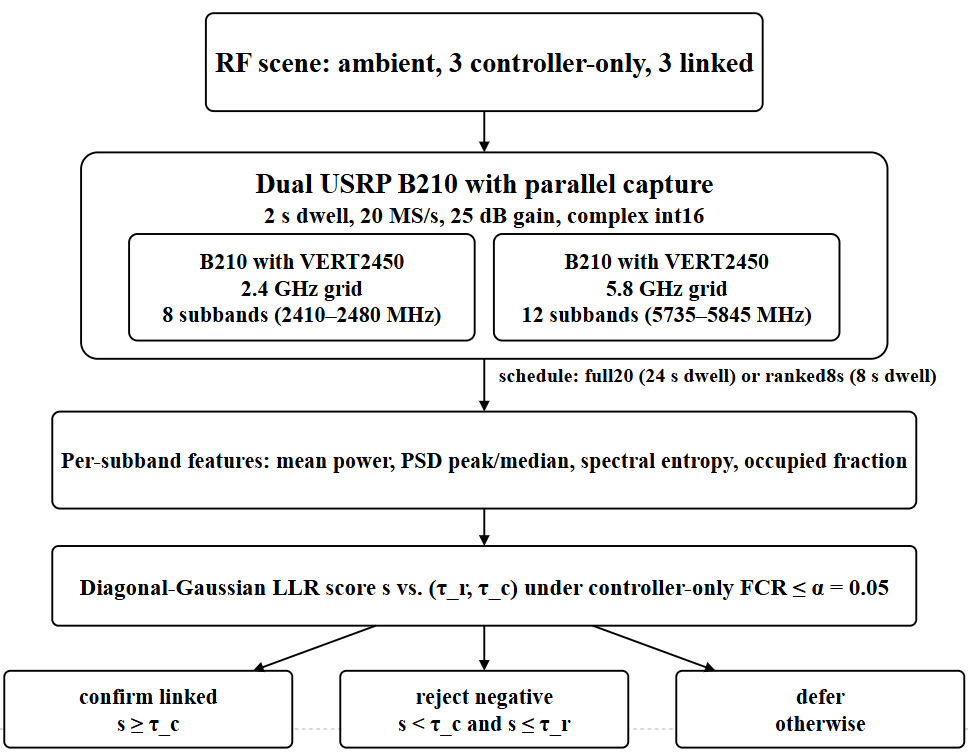}
    \caption{Dual-band passive RF linked-UAV confirmation pipeline.}
    \label{fig:system}
\end{figure}

Fig.~\ref{fig:system} summarizes the measurement pipeline along with methodology. Two Ettus Research USRP B210 receivers are connected to a host PC using USB 3.0. Each receiver uses a VERT2450 omnidirectional antenna. One USRP B210 scans the 2.4\,GHz ISM band using eight subbands at 2410 to 2480\,MHz in 10\,MHz steps. The other B210 scans the 5.8\,GHz ISM band using twelve subbands at 5735 to 5845\,MHz in 10\,MHz steps. The two receivers operate simultaneously to reduce the worst-case scan-cycle time from 40\,s with a single receiver covering all 20 subbands at 2\,s each to 24\,s, limited by the slower 5.8\,GHz receiver.

For each subband, the system captures complex int16 IQ samples at a sampling rate of 20 \, MS/s for 2 \,s with a receiver gain of 25 \, dB. A retry policy attempts each subband up to six times in total if UHD reports an overflow, underflow, or other runtime issue. Captures with no runtime overflow/underflow are used for reporting results in this paper (five clipping-amplitude warnings are retained and reported as a sensitivity case).

With this setup, eight 2.4\,GHz subbands cover approximately 2400--2490\,MHz and twelve 5.8\,GHz subbands cover approximately 5725--5855\,MHz. Neighboring captures overlap by 10\,MHz, and per-subband feature vectors at adjacent centers are spectrally correlated. The $2\times$ oversampling reduces sensitivity to passband-edge attenuation. All training, subband ranking, and evaluation use the same scan grid, so, the correlation across adjacent subbands is consistent between training and evaluation.

\subsection{Collection Environment}
\label{sseccollectionEnv}

\begin{figure}[h]
    \centering
    \includegraphics[width=\linewidth]{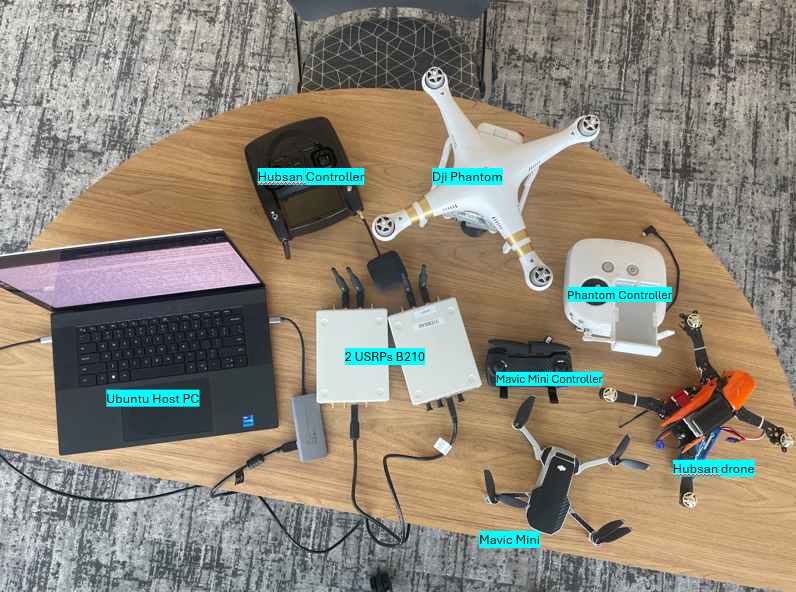}
    \caption{Experimental Setup.}
    \label{fig:experiment}
\end{figure}

RF signals were captured in three indoor rooms: a large laboratory with many pieces of equipment in place, a medium-sized classroom, and a small meeting room. The Wi-Fi access points were operational in all rooms and commercial cellular service was available throughout the building. The approximate distance between the UAV/controller equipment and the USRP antennas was 1.5\,m to 2.5\,m. The exact per-round separation and the locations and transmit powers of the ambient Wi-Fi access points were not recorded. The two B210 receivers were placed side-by-side. The orientation and position of the antenna were intentionally varied in all rounds to introduce diversity. Fig.~\ref{fig:experiment} shows an experimental setup.

\subsection{Platforms and States}
\label{ssec:plaform}

The three commercial UAV platforms are DJI Phantom~3 4K, Hubsan H501S, and DJI Mavic~Mini with the MR1SS5 controller. These platforms span different RF link architectures. Hubsan H501S uses 2.4\,GHz command and 5.8\,GHz FPV video. DJI Phantom~3 4K uses a 5.8 \, GHz remote control link and a 2.4 \, GHz Wi-Fi live-view/video path through DJI~GO. DJI Mavic~Mini with the MR1SS5 controller uses 5.8 \, GHz for remote control/video transmission, with the mobile device connected by cable.

The distinction in this paper is not split-band versus single-band operation. Instead, the important measurement question is whether the linked state produces additional RF evidence in subbands that are weak during the controller-only condition, or whether the linked and controller-only conditions overlap in the same subbands. In our measurements, Hubsan and Mavic Mini show additional linked-state evidence in subbands that are weak during controller-only captures. Phantom shows weaker linked-state separation in the 2.4\,GHz region where controller/phone Wi-Fi activity is already present. The main dataset (\texttt{r01-r20}) contains 20 rounds and seven RF states: ambient, controller-only, and linked for each platform. The linked-state IQ captures differ by platform and reflect each platform's normal live-view configuration.

\begin{itemize}
    \item \emph{Hubsan linked}: the controller's integrated display shows live FPV video during IQ capture.
    \item \emph{Mavic Mini linked}: a mobile device is connected to the controller by cable and the DJI Fly application displays live video during IQ capture.
    \item \emph{Phantom 3 linked}: In the \texttt{r01-r20} protocol, a smartphone is connected to the Phantom controller Wi-Fi and DJI~GO is used to confirm the linked state, but the phone is disconnected before IQ capture. In the \texttt{p01-p10} protocol, the phone remains connected to the controller Wi-Fi SSID and DJI~GO shows live video throughout every IQ capture.
\end{itemize}

Using both Phantom protocols allows us to measure the effect of phone-connected live-view configuration during capture. The phone-connected dataset separates \texttt{phantom\_linked\_phone\_connected} from \texttt{phantom\_controller\_only\_phone\_connected}. In the latter case, the controller is turned on and the phone remains connected, but the aircraft is off.

A separate hardware-in-loop (HIL) follow-up (rounds \texttt{h01-h05}) physically executes paired full and compact ranked schedules on the two USRPs for Hubsan and Mavic Mini, rather than estimating their durations from stored captures. Table~\ref{tab:dataset} summarizes the datasets.

\begin{table}[h]
\centering
\caption{Dataset Summary}
\label{tab:dataset}
\setlength{\tabcolsep}{2pt}
\renewcommand{\arraystretch}{1.02}
\begin{tabular}{@{}lrrr@{}}
\toprule
Dataset & Rounds & Scan-Cycles & IQ capture \\
\midrule
Main (r01--r20) & 20 & 140 & 2{,}800 \\
Phantom phone (p01--p10) & 10 & 20 & 400 \\
HIL timing (h01--h05) & 5 & 40 & 560 \\
\bottomrule
\end{tabular}
\end{table}

\section{Methodology}
\label{sec:method}

\subsection{Features}
\label{ssec:features}

For each 2\,s subband IQ capture, four per-subband features are extracted from the first 0.25\,s over the full sampled bandwidth without prior narrowband filtering. The features are: (1) mean received power in dB, (2) peak-to-median power spectral density, (3) spectral entropy, and (4) 6\,dB occupied fraction. The PSD is computed using Welch's method on the 0.25\,s segment with a 4096 sample Hann window and 50\% overlap. A scan cycle is represented by the concatenation of per-subband features across all observed subbands. The omitted-negative audit, controller-aware main-dataset comparison, and band ablation use mean received power and the Phantom phone-connected sensitivity analysis uses all four features.

\subsection{Scoring Model}
\label{ssec:scoring}

Per-subband features are standardized using training-round statistics. For each subband and feature, we fit diagonal-Gaussian class models for the linked and non-linked classes using training scan cycles. The non-linked class includes controller-only and ambient scans when those states are present in the evaluated subset. A log-likelihood-ratio (LLR) contribution is then computed for each observed subband. The scan-cycle score $s$ is the sum of the LLR contributions per-subband in the observed subbands.

\subsection{Omitted-Negative Audit}
\label{ssec:omittedAudit}

The omitted-negative audit represents a linked-versus-background evaluation. In each LORO fold, all controller-only scans were removed from feature normalization, Gaussian fitting, subband ranking, and threshold selection. The remaining training states are linked and ambient. A binary threshold is selected to maximize the balanced accuracy of the training linked-versus the ambient. Exact ties are resolved by a higher ambient true-negative rate, then linked TPR, then the higher threshold. The learned model and threshold are then applied unchanged to the held-out round's linked, ambient, and controller-only scans. Hence, controller-only data enter only as an audit set after the detector has been fixed.

This experiment does not implement any particular previously published classifier. It tests the narrower methodological question of whether a high linked-versus-ambient result can coexist with controller-only false confirmation in the same measurements.

\subsection{Controller-Only FCR-Constrained Decision Rule}
\label{ssec:controllerOnly}

For the controller-aware analyses, ambient and controller-only scans form the non-linked training class. We first evaluate an ordinary binary threshold selected to maximize training balanced accuracy, and then compare it with the constrained rule on the same subsets and folds. Our constrained decision rule uses two thresholds as defined below in Eq.~\ref{eq:decision}. The confirm
  threshold $\tau_c$ controls the cost of false confirmations on
  controller-only scans, and the reject threshold $\tau_r$ controls the
  cost of rejecting linked scans. Let $\fcrctrl$ denote the rate of
  \emph{confirm-linked} outputs on controller-only scan cycles. Given a
  tolerance $\alpha$, we select $\tau_c$ from training-data thresholds
  satisfying $\fcrctrl \leq \alpha$. Among feasible thresholds, we maximize
  linked TPR, using balanced accuracy and then negative TNR as tie-breakers. The threshold $\tau_r$
  is selected separately to maximize negative rejection, subject to
  linked-rejection rate $\leq \alpha$.
  
  Because the two searches are independent, $\tau_r$ can fall above or
  below $\tau_c$ depending on the training fold. The rule is therefore
  applied in a fixed order: confirm first, then reject.
  
  \begin{equation}
  d(s) =
  \begin{cases}
  \text{confirm linked}, & s \geq \tau_c,\\
  \text{reject negative}, & s < \tau_c \text{ and } s \leq \tau_r,\\
  \text{defer}, & \text{otherwise}.
  \end{cases}
  \label{eq:decision} 
  \end{equation}

  A defer is a score that is neither confirmed by the first rule nor rejected
  by the second. We set $\alpha = 0.05$ for both training constraints. This
  training-fold constraint does not guarantee that held-out $\fcrctrl$ is at
  most 0.05.

\subsection{Compact Ranked Scan}
\label{ssec:rankedScan}

A ranked subband order is learned from training rounds by sorting scan windows by their single-subband LLR separability. The compact dual-receiver scan is ranked separately within the 2.4\,GHz and 5.8\,GHz grids. The compact ranked scan (\texttt{ranked8s}) observes four dual-receiver dwell steps: four 2.4\,GHz scan windows and four 5.8\,GHz scan windows, for a nominal dwell time of $4 \times 2 = 8$\,s. The full scan (\texttt{full20}) observes all twenty scan windows, with a nominal dwell time of 24\,s because the 5.8\,GHz receiver requires twelve dwell steps.

For a full-window band ablation, the controller-aware analysis is retrained using all eight 2.4\,GHz windows, all twelve 5.8\,GHz windows, or all twenty windows.

\subsection{Evaluation}
\label{ssec:evaluation}

The main dataset is evaluated with leave-one-round-out (LORO) cross-validation. In each fold, one round is held out for testing, and the remaining 19 rounds are used for feature normalization, Gaussian model parameter estimation, threshold selection, and subband ranking. Bootstrap 95\% confidence intervals use 10{,}000 round-level resamples of the held-out predictions. The Phantom phone-connected follow-up is evaluated separately on rounds \texttt{p01-p10}, including a sensitivity analysis that excludes round \texttt{p01} due to clipping warnings. For HIL timing validation, models trained on the \texttt{r01}-\texttt{r20} data are applied to held-out \texttt{h01}-\texttt{h05} captures. When reporting HIL decision preservation, platform-specific models are used. Wall-clock time is measured from command issuance until both receiver processes end. A Hubsan+Mavic result pools separately collected scan cycles which means it is not a simultaneous two-platform experiment.

\section{Results}
\label{sec:results}

\subsection{Controller-Only Audit and Constrained Confirmation}
\label{per-Paltformconformation}

At four ranked dwell steps (8\,s), the omitted-negative detector confirms 59 out of 60 linked scans and rejects all 20 ambient scans. Its linked-versus-ambient balanced accuracy is therefore 0.992 [0.975, 1.000]. Without retraining, the same pooled all-platform detector falsely confirms 30 out of 60 controller-only scans, giving $\fcrctrl=0.500$ [0.400, 0.617]. These false confirmations comprise 19 out of 20 from Phantom, 6 out of 20 from Hubsan, and 2 out of 20 from Mavic controller-only scans. These values do not show that every detector trained on a public dataset will have the same error, but they show that a high linked-versus-ambient score did not reveal the controller-only behavior of this detector.

Table~\ref{tab:main}  compares thresholds trained and evaluated in identical subsets on the same four-step scan. For the all-platform model, the ordinary balanced-accuracy threshold gives the linked TPR 0.900 and $\fcrctrl=0.350$. The constrained rule reduces the remaining $\fcrctrl$ to 0.050, but the linked TPR drops to 0.400 and 59 of 140 scans are deferred. For Phantom, the same tradeoff is 0.700 TPR and 0.700 FCR at the ordinary threshold versus 0.250 TPR, 0.050 FCR, and 0.500 defer rate under the constrained rule. The pooled Hubsan+Mavic operating point is unchanged by the constraint. Consequently, this constraint is a risk-control mechanism rather than a uniformly successful separator. A lower observed FCR can be obtained by declining to confirm difficult scans, but the resulting TPR and defer rate must be reported with it.

\begin{table}[t]
\centering
\caption{Same-subset LORO threshold comparison at four ranked dwell steps (8\,s).}
\label{tab:main}
\setlength{\tabcolsep}{2.7pt}
\begin{tabular}{llrrrr}
\toprule
Subset & Rule & TPR & $\fcrctrl$ & $\fcramb$ & Defer \\
\midrule
All & Binary & 0.900 & 0.350 & 0.050 & 0.000 \\
All & Constrained & 0.400 & 0.050 & 0.000 & 0.421 \\
Hubsan+Mavic & Binary & 0.975 & 0.025 & 0.000 & 0.000 \\
Hubsan+Mavic & Constrained & 0.975 & 0.025 & 0.000 & 0.000 \\
Phantom & Binary & 0.700 & 0.700 & 0.050 & 0.000 \\
Phantom & Constrained & 0.250 & 0.050 & 0.000 & 0.500 \\
\bottomrule
\end{tabular}
\end{table}

Under the constrained rule, Hubsan separately reaches TPR 1.000 with zero observed FCR and deferral at four steps. Mavic reaches TPR 0.850 with zero FCR and 0.300 deferral at four steps. At six steps, TPR increases to 0.950 with zero FCR and deferral. Phantom is the least separable platform in the main dataset. At six steps its TPR is 0.250, FCR is 0.050, and defer rate is 0.550. The \texttt{r01}-\texttt{r20} Phantom captures used the no-phone-during-capture protocol, so Section~\ref{sec:phantom_phone} evaluates the phone-connected condition separately.

\subsection{Phantom Phone-Connected Protocol and Sensitivity}
\label{sec:phantom_phone}

The \texttt{p01}-\texttt{p10} Phantom follow-up keeps the phone connected to the controller Wi-Fi while DJI~GO displays live video throughout capture. We evaluate this follow-up separately using LORO over the ten follow-up rounds and the energy-spectral feature set. Table~\ref{tab:phantom_sensitivity} reports the full scan and an offline four-step restriction of those captures.

  \begin{table}[!h]
  \centering
  \caption{Phantom phone-connected follow-up sensitivity.}
  \label{tab:phantom_sensitivity}
  \setlength{\tabcolsep}{3.4pt}
  \begin{tabular}{llrrrr}
  \toprule
  Data & Policy & Steps & BA & TPR & $\fcrctrl$ \\
  \midrule
  p01--p10 & full20 & 12 & 0.700 & 0.700 & 0.300 \\
  p02--p10 & full20 & 12 & 0.778 & 0.667 & 0.111 \\
  p02--p10 & offline ranked8s & 4 & 0.444 & 0.444 & 0.556 \\
  \bottomrule
  \end{tabular}
  \end{table}

Excluding \texttt{p01} increases full20 balanced accuracy and reduces FCR, but compact early confirmation remains unreliable. Offline analysis of the full20 captures, restricted to four ranked dual-receiver dwell steps, gives controller-only FCR 0.556 after excluding \texttt{p01}. Thus, the Phantom result is not solely an artifact of \texttt{p01} clipping. It is consistent with overlap between controller/phone Wi-Fi activity and linked-state RF evidence in the 2.4\,GHz band, although these measurements do not isolate architecture from every other source of variation.

Under the phone-connected protocol, Phantom does produce discriminative 2.4\,GHz evidence, where the mean linked-minus-controller-only energy in the 2440--2470\,MHz range is $+1.6$ to $+3.8$\,dB. The corresponding separations for Hubsan and Mavic Mini are comparatively larger, i.e. $+14.8$\,dB at 5765\,MHz for Hubsan H501S and $+25.4$\,dB at 5755\,MHz for DJI Mavic~Mini. The phone-connected protocol therefore does not remove the controller-only confound in these measurements.

\subsection{Hardware-in-Loop Timing}
\label{ssec:hil_timing}

The Phase~B HIL experiment physically ran paired full20 and ranked8s scans for Hubsan and Mavic~Mini only. Wall-clock time measures elapsed time from command issuance until both receiver processes end. Table~\ref{tab:hil} shows that full20 averages 82.9\,s against a 24\,s nominal dwell time, while ranked8s averages 28.9\,s against 8\,s, a 65.2\% reduction. Non-dwell time includes startup, UHD initialization, retuning, USB transfer, file I/O, and any retries as these components were not timed separately. No physical ranked8s Phantom captures were collected.
\begin{table}[t]
  \centering
  \caption{Hardware-in-loop wall-clock timing. Platform rows pool ten trials; combined rows pool twenty.}
  \label{tab:hil}
  \begin{tabular}{lcrrr}
  \toprule
  Platform & Scan & Mean (s) & SD (s) & Range (s) \\
  \midrule
  Hubsan & full20 & 84.3 & 9.7 & 71--97 \\
  Hubsan & ranked8s & 29.0 & 5.7 & 24--37 \\
  Mavic~Mini & full20 & 81.4 & 7.5 & 71--97 \\
  Mavic~Mini & ranked8s & 28.8 & 4.9 & 23--37 \\
  Combined & full20 & 82.9 & 8.6 & 71--97 \\
  Combined & ranked8s & 28.9 & 5.2 & 23--37 \\
  \bottomrule
  \end{tabular}
  \end{table}
Under platform-specific constrained models, ranked8s preserves the full20 decision output in the HIL trials. For both Hubsan and Mavic~Mini, all five controller-only scans are rejected and all five linked scans are confirmed for both scan kinds, giving TPR = 1.000, $\fcrctrl = 0.000$, and zero deferrals. However, this does not hold for the all-platform threshold. Under the all-platform model trained on the full main dataset, Mavic linked scans remain confirmed, but Hubsan linked scans defer or reject under the more conservative threshold.

\subsection{Importance of Subbanding}
\label{ssec:subbandImportance}

The rankings recover expected operating regions. For Hubsan, the strongest 5.8\,GHz peaks are at 5765--5775\,MHz, consistent with the FPV video link, with secondary 2.4\,GHz peaks at 2420--2430\,MHz. For Mavic~Mini, the strongest peaks are at 5735--5755\,MHz, within the lower portion of its 5.8\,GHz operating range. Phantom shows weaker peaks, primarily in the upper 2.4\,GHz region. A full-window energy ablation gives TPR 1.000, zero observed controller-only FCR, and zero deferral for both Hubsan and Mavic using either 5.8\,GHz alone or simultaneous dual-band capture. This is limited to this dataset and model.

\section{Discussion and Limitations}
\label{sec:disc}

The omitted-negative audit provides the direct motivation that was absent from a linked-versus-background evaluation i.e. high balanced accuracy did not reveal the detector's response to a powered controller. The constrained comparison also shows the cost of controlling that response. Lower FCR is obtained partly by deferring uncertain observations. So TPR, FCR, and defer rate must be reported together.

Results vary substantially by platform. Hubsan and Mavic have strong 5.8\,GHz separation in this dataset, whereas Phantom's weaker phone-connected evidence overlaps 2.4\,GHz controller/phone activity. With only three products, these are platform-dependent measurements rather than a general architectural taxonomy.

\subsection{Limitations}
Our study has the following limitations:
  \begin{itemize}
      \item \emph{Platform scope:} Only three commercial UAV platforms are evaluated. Generalization to other COTS or military platforms, especially platforms that use bands outside the scanned ISM grids, needs further extensions.

      \item \emph{Environment:} RF data capturing was done indoor at approximately 1.5\,m to 2.5\,m distance, need to be extended outdoors with varying ranges.

  \end{itemize}
Future experiments should test whether the detector still works when distance and environmental conditions change. Similarly, it needs evaluation of more UAV models, and use features beyond simple signal power when the controller and linked aircraft occupy the same frequencies. Independent validation also requires datasets that explicitly distinguish a powered controller with no aircraft from a controller linked to an aircraft under comparable measurable conditions.

\section{Conclusion}
\label{sec:conc}

This work emphasizes that controller-only false confirmation should be reported separately from linked-versus-background accuracy. In the omitted-negative audit, a detector reaches 0.992 balanced accuracy while false-confirming 30 of 60 held-out controller-only scans. Including controller-only scans in training and applying the FCR constraint reduces the all-platform observed FCR from 0.350 to 0.050, but linked TPR falls from 0.900 to 0.400 and 42.1\% of scans are deferred. Results vary by platform: Hubsan and Mavic are separable at the selected operating points, whereas Phantom remains difficult under both capture protocols. Hardware-in-loop timing further shows that the compact ranked scan reduces measured scan time by 65.2\% for Hubsan and Mavic. These findings establish the controller-only omission for this dataset and detector, not universal behavior across platforms or distances.

\section*{Acknowledgment}
\label{sec:ack}
The authors thank Cybastion Technology
for their funding and support that made this research work
possible.

\bibliographystyle{IEEEtran}
\bibliography{refs}

\end{document}